\documentclass[12pt,preprint]{aastex}

\usepackage{graphicx,xcolor,amsmath}

\begin{document}

%\preprint{}

\title{Physical role of topological constraints in localised magnetic relaxation}


\author{A.~R. Yeates}
\email{anthony.yeates@durham.ac.uk}
%\thanks{}
\affil{Department of Mathematical Sciences, Durham University, Durham, DH1 3LE, United Kingdom}
\author{A.~J.~B. Russell}
\author{G. Hornig}
\affil{Division of Mathematics, University of Dundee, Dundee, DD1 4HN, United Kingdom}


%\date{\today}

\begin{abstract}
Predicting the final state of turbulent plasma relaxation is an important challenge, both in astrophysical plasmas such as the Sun's corona and in controlled thermonuclear fusion. Recent numerical simulations of plasma relaxation with braided magnetic fields identified the possibility of a novel constraint, arising from the topological degree of the magnetic field-line mapping. This constraint implies that the final relaxed state is drastically different for an initial configuration with topological degree 1 (which allows a Taylor relaxation) and one with degree 2 (which does not reach a Taylor state). Here we test this transition in numerical resistive-magnetohydrodynamic simulations, by embedding a braided magnetic field in a linear force-free background. Varying the background force-free field parameter generates a sequence of initial conditions with a transition between topological degree 1 and 2. For degree 1, the relaxation produces a single twisted flux tube, while for degree 2 we obtain two flux tubes. For predicting the exact point of transition, it is not the topological degree of the whole domain that is relevant, but only that of the turbulent region.
\end{abstract}

%\pacs{52.30.Cv, 52.35.Vd, 52.65.Kj, 96.60.Hv}
%\keywords{}


\maketitle

\section{Introduction}

Self-organisation of turbulently relaxing plasma to a predictable minimum-energy state has been observed in laboratory confinement devices including the reversed-field pinch and the spheromak \cite{1974PhRvL..33.1139T,1986RvMP...58..741T, ortolani, 2004RvMP...76.1071B}. The so-called Taylor relaxation hypothesis assumes that the only relevant constraints on the dissipation of magnetic energy are the total magnetic flux and the total magnetic helicity. The latter is not an exact invariant in the presence of resistivity, but is known to be well-preserved on typical timescales of relaxation processes. In order that all other ideal invariants are destroyed (such as helicity in subregions of the plasma \cite{1986RvMP...58..741T}, or other helicity moments \cite{1980PhRvL..45..347B}), the evolution must be sufficiently turbulent that magnetic reconnection is able to occur throughout the volume.

It has also been proposed that this Taylor relaxation theory might be applied to predict the energy released by rapid heating events in the solar corona \cite{1984A&A...137...63H}, where magnetic energy is believed to be released through relaxation to a lower-energy equilibrium. In this context, numerical magnetohydrodynamic (MHD) simulations have modelled the dynamic relaxation of various initially unstable equilibria, such as kink-unstable twisted magnetic flux ropes \cite{1996A&A...308..935B,1999ApJ...517..990A,2008A&A...485..837B, 2011SoPh..273...93B,2013A&A...550A..40B}, or a magnetic field with a braided structure \cite{2011A&A...525A..57P, 2011A&A...536A..67W}. Our work has been motivated by the latter simulations, which showed that certain initial configurations self-organised into final equilibria whose magnetic topology was more complicated than predicted by Taylor theory, despite the occurrence of efficient reconnection. We identified the presence of an additional constraint beyond the total magnetic flux and helicity: the topological degree of the field line mapping \cite{2010PhRvL.105h5002Y, 2011JPhA...44z5501Y}.

The topological degree (defined in Section \ref{sec:degree}) is conserved provided that the degree of the boundary does not change. The latter can be ensured by having turbulent dynamics that are localised in the interior of the domain and do not affect the boundary. It is our goal in this paper to show, for a sequence of initial conditions of degree 1 which approach degree 2, how the final state suddenly switches from a single flux tube to a pair of flux tubes. 

The assumption of localisation is an important one for relaxation events in the solar corona. Unlike the reversed-field pinch, there are no conducting walls to define
the relaxation volume \cite{2013SSRv..178..325B}. Typically, coronal energy releases--for example, in solar flares--are highly localised in space. The extent of the relaxation region is determined by the connectivity of the magnetic field configuration, requiring either unstable configurations or very small-scale gradients to initiate the energy release. Dixon \textit{et al.} \cite{1989A&A...225..156D} showed that Taylor theory may be applied to regions with a free boundary, although they did not specify where the boundary should be placed in any particular magnetic field. More recently Bareford \textit{et al.} \cite{2013A&A...550A..40B} have shown that Taylor theory can give reasonable predictions of relaxed states in numerical solutions of kink-unstable magnetic flux tubes, provided that it is applied within the appropriate subregion.

Localised Taylor relaxation has also been applied in the context of tokamaks. In these devices, global Taylor relaxation does not describe the magnetic configurations that are observed. However, Hudson \textit{et al.} \cite{2012PPCF...54a4005H} have developed a partial relaxation model where Taylor relaxation occurs in sub-volumes. These sub-volumes are separated by a discrete set of irrational flux surfaces that survive even in the presence of the chaotic field lines typical of non-axisymmetric magnetic fields. In another application, Gimblett \textit{et al.} \cite{2006PhRvL..96c5006G} have developed a model for edge-localised modes based on localised Taylor relaxation within only the outer region of the plasma.

In this paper, we consider a one-parameter family of initial magnetic configurations in a periodic (topologically toroidal) domain. These configurations, described in Section \ref{sec:init}, are chosen to have a ``background field'' of gradually varying structure.  This complements the particular configurations where this constraint was demonstrated previously \cite{2010PhRvL.105h5002Y}, which had vanishing magnetic helicity.
 
\section{Topological degree constraint} \label{sec:degree}

\begin{figure}
\centering
\includegraphics[width=0.7\textwidth]{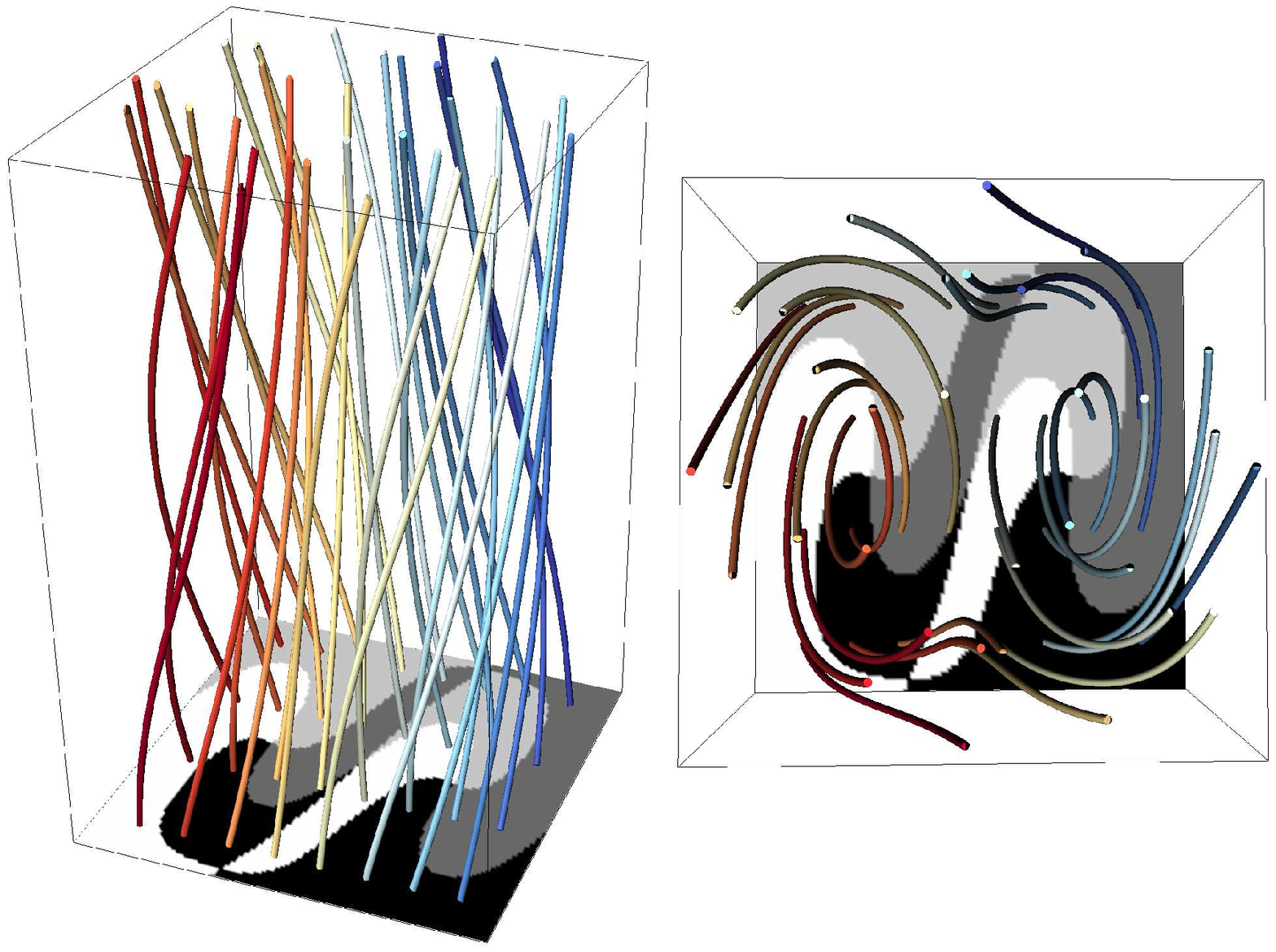}
\caption{The magnetic field given by \eqref{eqn:eg}, showing the colour map (in grayscale) and selected magnetic field lines. There are three fixed points with $T(D)=1$ for the region shown. \label{fig:degree}}
\end{figure}

To define the topological degree of a particular configuration, let $f:D_0\to D_1$, where $f=(f_x,f_y)$, be the field line mapping from the lower boundary $D_0$ to the upper boundary $D_1$. In other words, $f({\bf x}_0)\in D_1$ is the end-point of the magnetic field line starting at ${\bf x}_0\in D_0$. We assume that there is a strong enough guide field that all field lines pass from $D_0$ to $D_1$ without changing direction. We shall assume for simplicity that $D_1=D_0$, as in the periodic simulations presented in this paper. Field lines that satisfy $f({\bf x}_0)$ = ${\bf x}_0$ are known as fixed points of $f$ (or periodic orbits in the case of periodic boundaries).  The index of a fixed point describes the local structure of $f$ around the fixed point, and is defined as the local Brouwer degree of $f$ (for more details, see Yeates \& Hornig \cite{2011JPhA...44z5501Y}). Now let $D \subset D_0$ be a subregion of $D_0$. The topological degree of $f$ on $D$, denoted $T(D)$, is defined to be the total (net) fixed point index, obtained by summing the indices of all isolated fixed points of $f$ in $D$. One may express $T(D)$ as the Kronecker integral
\begin{equation}
T(D) = \frac{1}{2\pi}\oint_{\partial D}{\rm d}\left[\arctan\left(\frac{f_y - y}{f_x - x}\right) \right],
\end{equation}
around the boundary of $D$ \cite{2003Chaos..13...94P}. Since $T(D)$ is an integer, the only way it can change under a continuous time-evolution of $f$ is if one or more fixed points cross into or out of the boundary of $D$. So if $f$ is fixed on the boundary of our turbulent region $D$, then $T(D)$ must be preserved in time. In particular, this means that the relaxed state may be forced to contain more than one fixed point, implying certain magnetic substructure.

We utilize the convenient colour map technique introduced by Polymilis et al. \cite{2003Chaos..13...94P} for visualizing fixed points of $f$, their indices, and $T(D)$. This is illustrated in Figure \ref{fig:degree} with the magnetic field
\begin{eqnarray}
{\bf B} &=& \nabla\times{A{\bf e}_z} + {\bf e}_z,\nonumber\\
A &=& 0.6\sin^2x\cos(0.5y) + \cos(0.3x)\cos(0.3y).
\label{eqn:eg}
\end{eqnarray}
The colour map assigns one of four colours (in this paper, we use shades of gray) to each point $(x,y)$ in $D_0$, according to the relative signs of $f_x - x$ and $f_y - y$. Fixed points are readily identified as places where all four colours intersect. Furthermore, the topological degree $T(D)$ of a region $D\subset D_0$ may be identified by noting the anticlockwise sequence of colours around the boundary of $D$. In particular, the number of times that the full sequence of four colours (in the correct order) is repeated. For example, the degree of the full region shown in Figure \ref{fig:degree} is $+1$. Correspondingly, there is a net anticlockwise rotation of field lines around the boundary. Inside $D$, there are three fixed points: two ``elliptic'' points with degree $+1$ and one ``hyperbolic'' point in the centre with degree $-1$.

The topological degree relates the complexity of the field on the boundary of the domain to that of the interior field. This is similar to how Gauss' theorem relates the integrated electric field over a closed surface to the electric charge inside the surface. As for the topological degree, the surface integral over the electric field does not distinguish how many positive or negative electric charges are inside the domain: it only gives a net charge. For the topological degree, the analogue of the net charge is the sum of hyperbolic (degree $-1$) and elliptic (degree $+1$) periodic orbits. The simplest state (the smallest number of charges which give the correct net charge) is typically also the one with lowest energy. Thus an efficient turbulent relaxation within an otherwise ideal plasma is expected to lead to the simplest force-free field consistent with the topological degree of the turbulent region.

\section{Numerical setup}

\subsection{Starting configurations} \label{sec:init}

In this paper, we present resistive-MHD simulations for a family of initial magnetic configurations. Each is a superposition of two components ${\bf B}={\bf B}_\alpha + {\bf B}_{\rm braid}$, where ${\bf B}_\alpha$ is a linear force-free field with constant $\alpha$, and ${\bf B}_{\rm braid}$ is a braiding magnetic field pattern consisting of six toroidal rings of magnetic flux. The field ${\bf B}_{\rm braid}$ is orthogonal to ${\bf e}_z$ and vanishes on the boundaries of our domain. By contrast, the background field ${\bf B}_\alpha$ is non-zero on all six boundaries of our domain. (For numerical convenience, we use a Cartesian domain.) By varying $\alpha$ and keeping ${\bf B}_{\rm braid}$ fixed, we obtain a one-parameter family of initial configurations. For $\alpha=0$ the configuration has degree 2, while for all positive values of $\alpha$ it has degree 1. By decreasing the value of $\alpha$ towards 0, we can test when and how the transition affects the relaxed state. To initialise the other variables in our resistive-MHD simulations, we simply take zero initial velocity, constant density and constant pressure. 

Note that the combined field ${\bf B}$ is not in equilibrium, and leads to a dynamical evolution in the resistive-MHD equations. Previous simulations (in the $\alpha=0$ case) have found consistent final-state topology whether or not the field is first subjected to an ideal relaxation before initiating the resistive-MHD evolution \cite{2010PhRvL.105h5002Y}.

For ${\bf B}_\alpha$, we take the well-known axisymmetric constant-$\alpha$ magnetic field of Lundquist \cite{1951PhRv...83..307L}. In cylindrical coordinates $(r,\phi,z)$, this takes the form
\begin{equation}
{\bf B}_\alpha = B_0\big(J_1(\alpha r){\bf e}_\phi + J_0(\alpha r){\bf e}_z\big),
\end{equation}
where $J_0$ and $J_1$ are Bessel functions of the first kind. This field is readily shown to satisfy $\nabla\times{\bf B}_\alpha=\alpha{\bf B}_\alpha$ for constant $\alpha$. In the limit $\alpha\to 0$, it reduces to a vertical, current-free magnetic field ${\bf B}_0=B_0{\bf e}_z$. In this paper we fix $B_0=1$. The condition that $B_z>0$ everywhere in our domain puts an upper limit on the acceptable values of $\alpha$. In particular, we require $\alpha<\alpha_{\rm r}$, where $\alpha_{\rm r}\approx 0.21$ is the smallest root of $J_0(\alpha_r \sqrt{128})=0$. (This is when the first field reversal occurs at the corners of the domain.) It should also be noted that ${\bf B}_\alpha$ leads to a net electric current in the $z$-direction.

The braiding field ${\bf B}_{\rm braid}$ was introduced by \cite{2009ApJ...696.1339W}; its construction is based on the pigtail braid, with six toroidal rings of flux,
%\begin{eqnarray}
%{\bf B}_{\rm braid} &=& \sqrt{2}\sum_{i=1}^6 k_i\exp\left(\frac{-(x-x_i)^2 - y^2}{2} - \frac{(z-z_i)^2}{4} \right) \nonumber\\
%&&\times\left(-y{\bf e}_x + (x-x_i){\bf e}_y\right).
%\end{eqnarray}
\begin{equation}
{\bf B}_{\rm braid} = \sqrt{2}\sum_{i=1}^6 k_i\exp\left(\frac{-(x-x_i)^2 - y^2}{2} - \frac{(z-z_i)^2}{4} \right)\Big(-y{\bf e}_x + (x-x_i){\bf e}_y\Big).
\end{equation}
The parameters used are $x_i=k_i=(1,-1,1,-1,1,-1)$, $z_i=(-20,-12,-4,4,12,20)$. This pattern of flux is efficient at ``mixing'' the field lines while having zero net helicity, and leads to a demonstrably chaotic field line mapping in our periodic domain \cite{2011A&A...536A..67W}. It is effectively this region of efficient mixing that generates small magnetic scales enabling current sheets to form, leading to magnetic reconnection. The extent of this region determines the region of turbulent relaxation in which the field is able to relax efficiently.

\begin{figure}
\centering
\includegraphics[width=\textwidth]{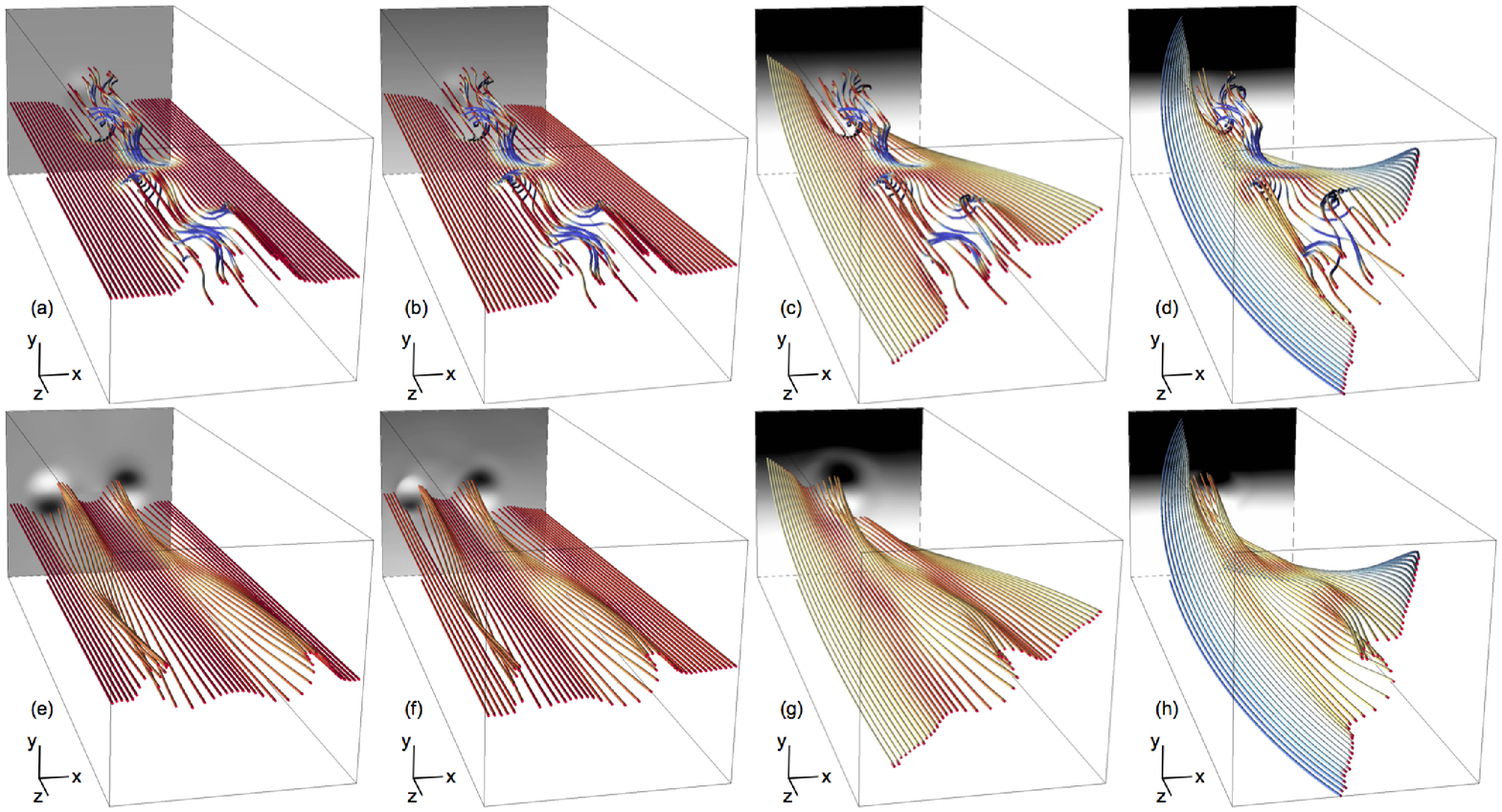}
\caption{Magnetic field lines for the simulations with $\alpha=0.001$, $0.01$, $0.05$ and $0.1$, at $t=0$ (a-d) and $t=300$ (e-h). The field lines are traced from a straight line on the mid-plane $z=0$, and are coloured by $B_\perp\equiv(B_x^2 + B_y^2)^{1/2}$ (with red for $0$ and saturated at $0.4$ in blue). The grayscale cross-section shows $B_x$ on the boundary $z=-24$ (black negative, white positive). \label{fig:lines}}
\end{figure}

Figure \ref{fig:lines} shows illustrative magnetic field lines for the combined states with $\alpha=0.001$, $0.01$,  $0.05$, and $0.1$. Although the field line connectivity is significantly altered, ${\bf B}_{\rm braid}$ is, energetically, a relatively small perturbation to the background field ${\bf B}_\alpha$. Denoting the magnetic energy by $E_{\rm mag}=\langle B^2\rangle/(2\mu_0)$, one has that $E_{\rm mag}({\bf B}_0 + {\bf B}_{\rm braid})\approx 1.008E_{\rm mag}({\bf B}_{0})$, while $E_{\rm mag}({\bf B}_{0.1} + {\bf B}_{\rm braid})\approx 1.009E_{\rm mag}({\bf B}_{0.1})$. It should be noted that ${\bf B}_\alpha$ is \emph{not} the minimum-energy (Taylor) state for our configuration, except when $\alpha=0$. This is because the magnetic helicity of the combined field ${\bf B}$ differs from that of ${\bf B}_{\alpha}$. 

\subsection{Numerical simulations}

The Lare3D Lagrangian-remap code \footnote{Lare3D is available from http://ccpforge.cse.rl.ac.uk/gf/.} is used to solve the resistive-MHD equations in a Cartesian box $\{-8\leq x\leq 8, -8\leq y\leq 8, -24\leq z\leq 24 \}$, at resolution $320\times 320\times 240$. We apply periodic boundary conditions in $z$ and line-tied boundary conditions in $x$ and $y$.  The code solves the non-dimensionalised equations
\begin{eqnarray}
\frac{\partial\rho}{\partial t} &=& -\nabla\cdot(\rho{\bf v}),\\
\rho\frac{D{\bf v}}{Dt} &=& {\bf j}\times{\bf B} - \nabla p + \nabla\boldsymbol{\sigma},\label{eqn:dv}\\
\frac{\partial{\bf B}}{\partial t} &=& \nabla\times({\bf v}\times{\bf B}) - \nabla\times(\eta{\bf j}),\\
\rho\frac{D\epsilon}{Dt} &=& -p\nabla\cdot{\bf v} + \eta j^2 + \boldsymbol\varepsilon\boldsymbol\sigma,\label{eqn:eps}\\
p &=& \rho \epsilon (\gamma -1),\\
\mu_0{\bf j} &=& \nabla\times{\bf B}.
\end{eqnarray}
Here $\rho$ is the mass density, ${\bf v}$ the plasma velocity, ${\bf B}$ the magnetic field, ${\bf j}$ the current density, $p$ the plasma pressure, $\boldsymbol\sigma$ the stress tensor, $\epsilon$ the specific internal energy density, $\eta$ the resistivity, $\boldsymbol\varepsilon$ the strain tensor, and $\gamma=5/3$ the ratio of specific heats. Details of the numerical methods are given by Arber \textit{et al.} \cite{2001JCoPh.171..151A}. The viscous term $\nabla\boldsymbol\sigma$ in \eqref{eqn:dv} includes no background viscosity, but only a shock viscosity to prevent unphysical oscillations and approximate the jump in entropy across shocks. The shock viscosity takes the tensor form given in \cite{2013A&A...550A..40B}, and we use the same parameter values $\nu_1=0.1$, $\nu_2=0.5$. There is a corresponding heating term $\boldsymbol\varepsilon\boldsymbol\sigma$ in \eqref{eqn:eps}.  We initially set $\rho=1$ and $\epsilon=0.01$ in non-dimensional units. In these units, one unit of time is equal to the time taken by an Alfv\'en wave with $B=\rho=1$ to move a unit distance in our box. The simulations presented here use a uniform resistivity of $\eta= 5\times 10^{-4}$. Previous simulations of the $\alpha=0$ case found that the topology of the relaxed state is not sensitive to the choice of $\eta$, although the details of the turbulent relaxation do change \cite{2011A&A...525A..57P}.

\section{Results}

For all values of $\alpha$, there is an initial phase of turbulent relaxation until approximately $t=100$, followed by a more gradual resistive dissipation. This pattern is the same as the earlier simulations with $\alpha=0$ \cite{2011A&A...525A..57P}, and was also seen for the relaxation of a kink-unstable loop \cite{2013A&A...550A..40B}. Huang \textit{et al.} \cite{2014ApJ...793..106H} find a similar distinction between quasi-static resistive evolution and the onset of a dynamical phase, in resistive reduced-MHD simulations of a randomly structured field.

\begin{figure}
\centering
\includegraphics[width=\textwidth]{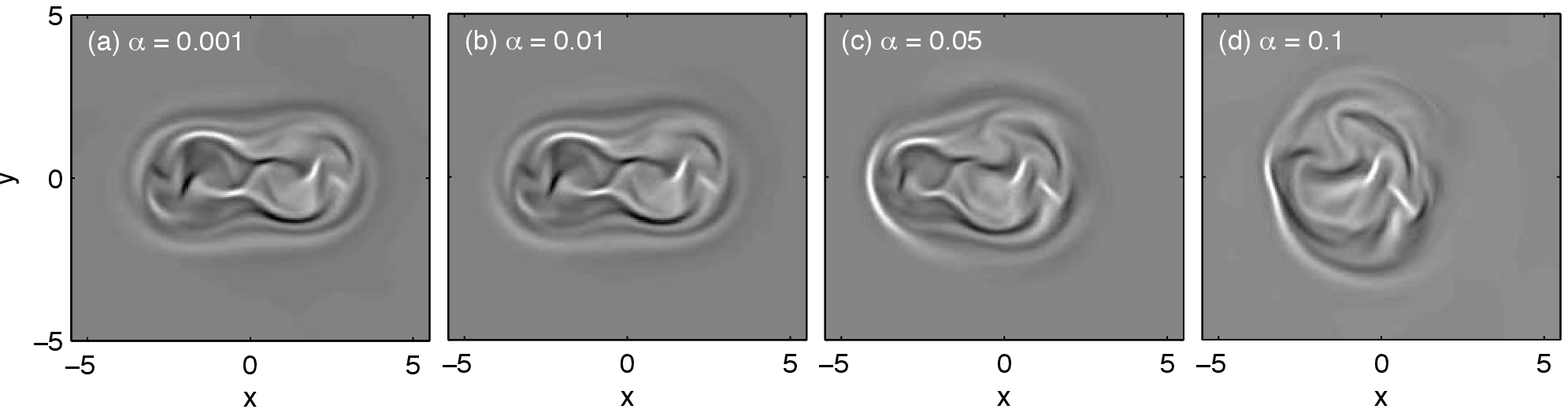}
\caption{Vertical current density $j_z$ in the mid-plane $z=0$ at $t=50$, for the simulations with $\alpha=0.001$, $0.01$, $0.05$, and $0.1$. The gray scale is saturated at $j_z=\pm 1$. \label{fig:jz}}
\end{figure}

\begin{figure}
\centering
\includegraphics[width=0.7\textwidth]{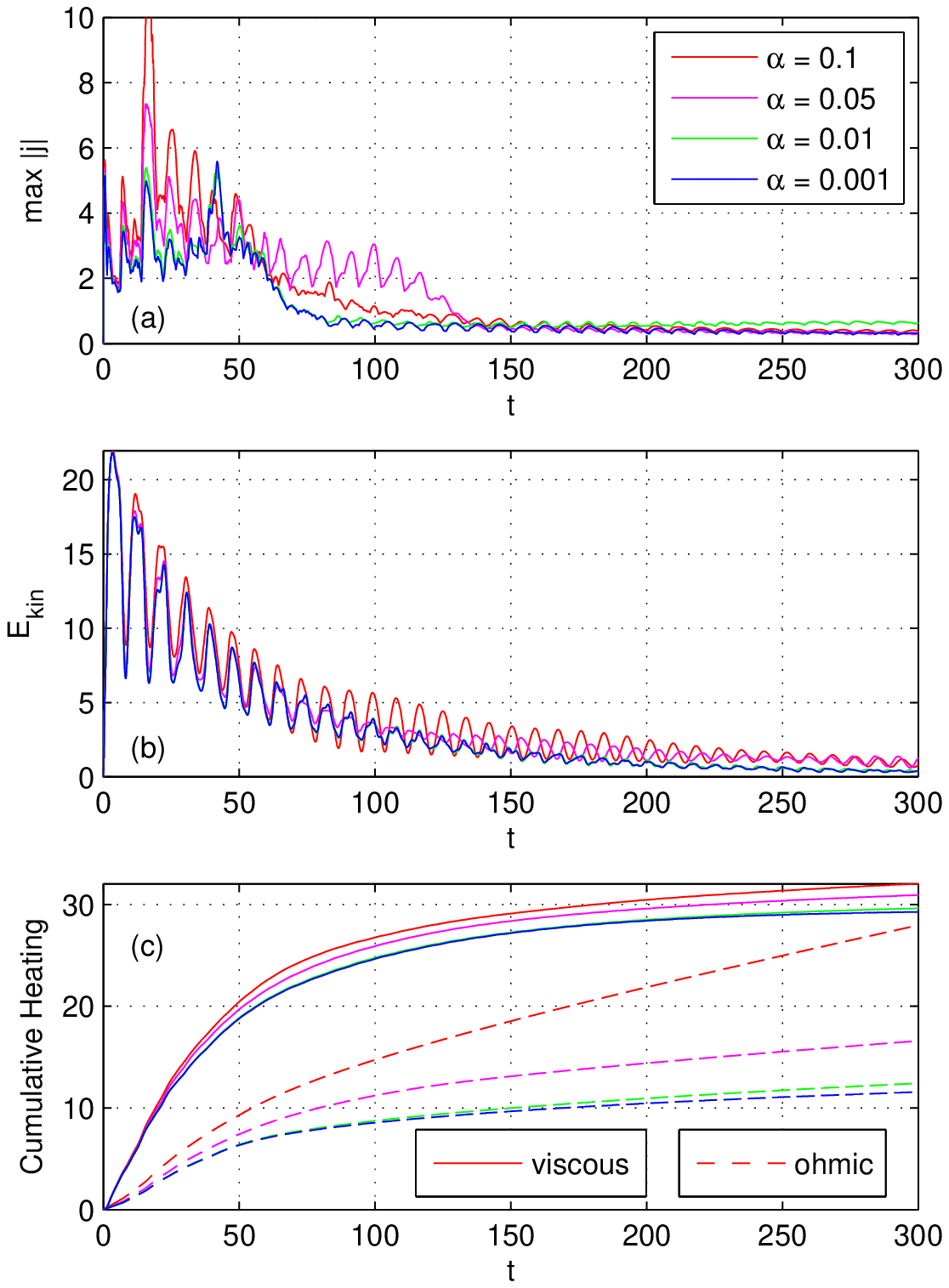}
\caption{Time evolution of (a) maximum current density $|{\bf j}|$, (b) total kinetic energy $E_{\rm kin}$, and (c) cumulative viscous (solid lines) and ohmic (dashed lines) heating, in the simulations with $\alpha=0.1$, $0.05$, $0.01$, and $0.001$.
\label{fig:energy}}
\end{figure}

In the turbulent phase of our simulations, the dynamics consists of a cascade from initially large to smaller current sheets, which interact with one another to dissipate magnetic energy during the relaxation. Figure \ref{fig:jz} shows the appearance of these current sheets at $t=50$ during the turbulent phase, in a cross-section at the mid-plane $z=0$. In each case there is a distinguished turbulent region outside which there are no significant currents or dynamics. The shape of the turbulent region is more circular for the run with $\alpha=0.1$, owing to the influence of the different background field. Figure \ref{fig:energy}(a) shows the maximum current throughout the domain as a function of time. All four runs follow a bursty, intermittent pattern of maximum current in the turbulent phase, followed by a smooth evolution with lower maximum current during the gradual, resistive phase. The run with $\alpha=0.05$ maintains a higher maximum current for longer than the others: this is due to the interaction of one of the resulting flux tubes with the background field, as will be discussed below. 

\begin{figure}
\centering
\includegraphics[width=0.7\textwidth]{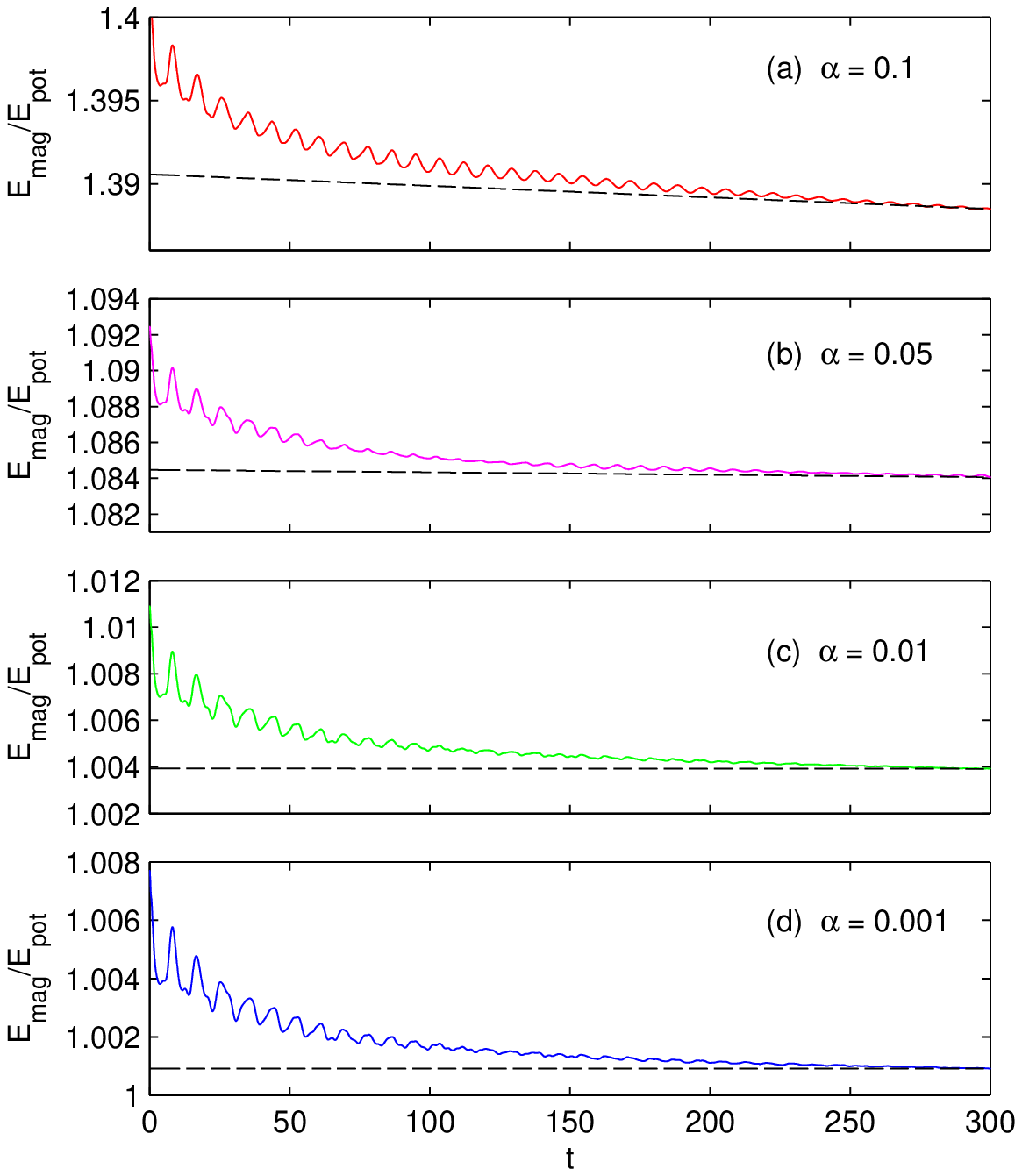}
\caption{Magnetic energy as a function of time in the simulations with (a) $\alpha=0.1$, (b) $0.05$, (c) $0.01$, and (d) $0.001$. Shown is the ratio of total magnetic energy to $E_{\rm pot}$ for each value of $\alpha$ (see text). The dashed lines show resistive decay rates $\exp(-\alpha^2\eta t)$ of the corresponding ${\bf B}_{\alpha}$ fields. 
\label{fig:enb}}
\end{figure}

The turbulent phase is also evident in the total energies shown in Figures \ref{fig:energy} and \ref{fig:enb}. For example, the total kinetic energy $E_{\rm kin}=\langle  \rho v^2\rangle/2$ is significant mainly during the turbulent phase, and follows a quite similar pattern in all runs. The oscillations seen in $E_{\rm kin}$ and also in the magnetic energy $E_{\rm mag}$ have a period consistent with torsional Alfv\'en waves, launched from the initial flux ring locations and counter-propagating in $z$. Although these waves dominate the frequency spectrum, the dynamics are nonetheless turbulent in the sense that the chaotic field line mapping produces a cascade to smaller spatial scales throughout the braided region. This cascade, and the resulting unpredictable bursts of reconnection, are important for removing energy and restructuring the magnetic field. Previous simulations of the $\alpha=0$ case have shown shown that consistent relaxed states are obtained whether the initial state contains discrete flux rings (as here), or is first subjected to an ideal relaxation \cite{2011A&A...525A..57P}, in which case there is a broader frequency spectrum.

In our resistive simulations, the dissipation of magnetic energy must be compared to that of the background ${\bf B}_\alpha$ field under resistive diffusion alone. Figure \ref{fig:enb} shows that the turbulent phase is characterized by a much faster dissipation of magnetic energy than would be expected from diffusion of ${\bf B}_\alpha$ (dashed line). In these plots, the energy is normalised by $E_{\rm pot}$, which is the energy of a uniform vertical field ${\bf B}=B_0{\bf e}_z$ with $B_0$ chosen to give the same magnetic flux as ${\bf B}_\alpha$. This is the minimum possible energy for each configuration in our periodic domain, ignoring all helicity constraints (and also the constraint of line-tying on the side boundaries). Some of the magnetic energy is lost by ohmic dissipation, but the majority of magnetic energy is dissipated by viscous heating at shock fronts, generated by the turbulent reconnection \cite{2013A&A...550A..40B}. During the turbulent phase, the rate of ohmic heating is only $20-50\%$ that of viscous heating. This is evident in Figure \ref{fig:energy}(c), which shows the cumulative viscous and ohmic heating in each run. Both heating rates level off after the end of the turbulent phase, although the asymptotic ohmic heating rates depend on $\alpha$, reflecting the resistive decay of the background field (which decays like $\exp(-\alpha^2\eta t)$ since it is a linear force-free field).

\begin{figure}
\centering
\includegraphics[width=\textwidth]{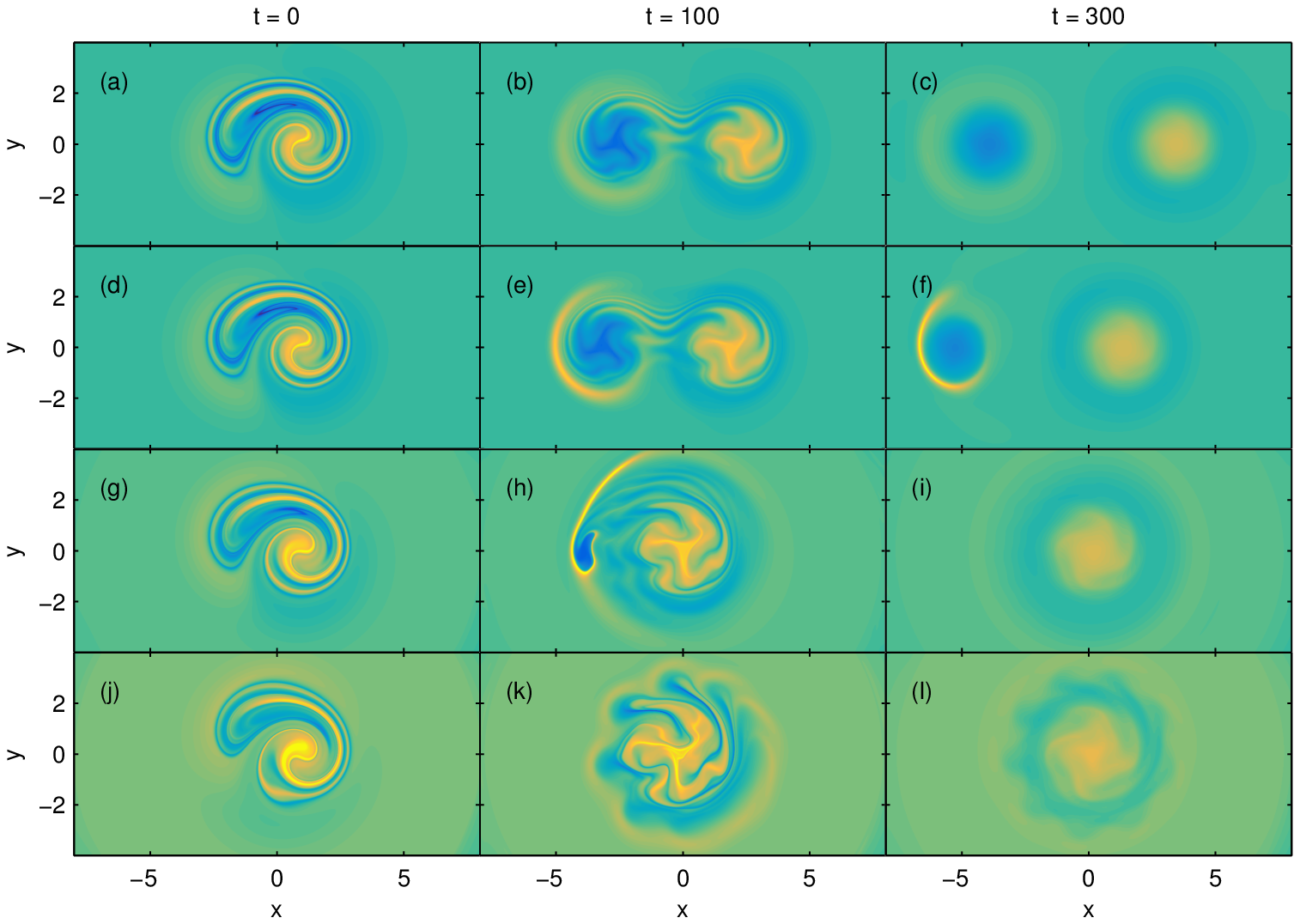}
\caption{The quantity $\bar{\lambda}$ for the simulations with $\alpha=0.001$ (a-c), $\alpha=0.01$ (d-f), $\alpha=0.05$ (g-i), and $\alpha=0.1$ (j-l). The left column shows $t=0$, the middle column shows $t=100$, and the right column shows $t=300$. The colour scale (blue/dark negative, yellow/light positive) is saturated at $\bar{\lambda}=\pm 0.5$. \label{fig:alpha}}
\end{figure}

In this paper, our main focus is on the magnetic topology of the end states. Here ``end state'' means the gradually decaying configuration that remains after the turbulent phase has ended. It is evident from the magnetic field lines at $t=300$ (Figure \ref{fig:lines}, bottom) that there is a difference between the end states for $\alpha=0.001$ and $\alpha=0.01$, as compared to $\alpha=0.05$ and $\alpha=0.1$. In the former two runs, there are two oppositely twisted flux tubes, while in the latter two runs there is only a single flux tube. We remark that Parker \cite{1983ApJ...264..635P} proposed that static MHD equilibria can not have more complicated topology than either a single flux tube or two parallel flux tubes of opposite twist, and our end states are in accord with this.

The separation into either one or two tubes is clearly seen in Figure \ref{fig:alpha}, which shows the average value $\bar\lambda$ of $\lambda={\bf j}\cdot{\bf B}/B^2$ along each magnetic field line. The quantity $\lambda$ is the current helicity density (we avoid the symbol $\alpha$ which refers specifically to the background field ${\bf B}_\alpha$). In a force-free equilibrium, which approximately holds after the turbulent relaxation, $\lambda$ is constant along each field line. Note that the separation into two tubes is not merely a transient phenomenon: the two twisted tubes for $\alpha=0.001$ or $\alpha=0.01$ actually repel one another and will not eventually merge together. Rather, their currents (twist) will continue to individually decay on the resistive timescale.  

\begin{figure}
\centering
\includegraphics[width=\textwidth]{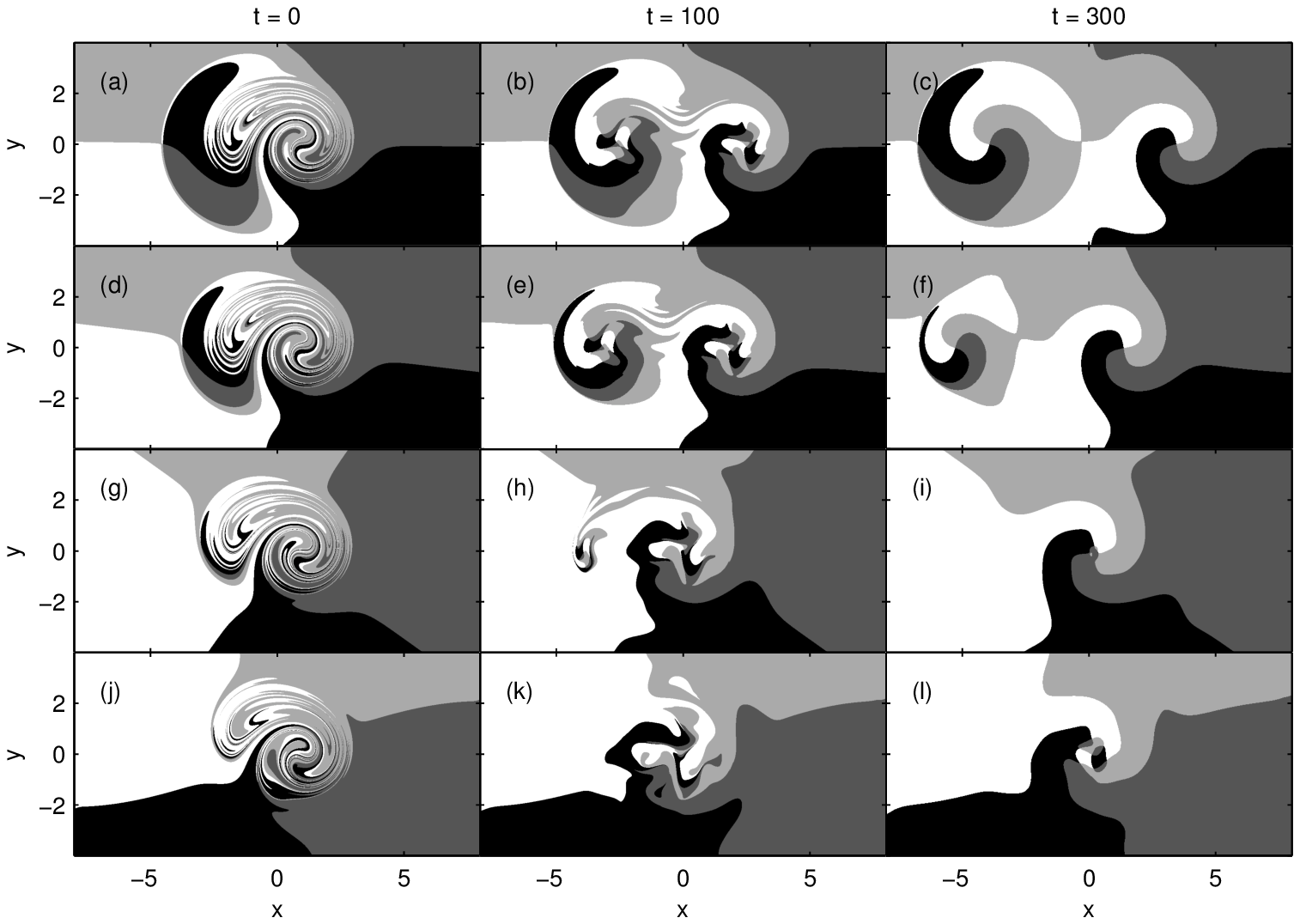}
\caption{Colour maps for the simulations with $\alpha=0.001$ (a-c), $\alpha=0.01$ (d-f), $\alpha=0.05$ (g-i), and $\alpha=0.1$ (j-l). The left column shows $t=0$, the middle column shows $t=100$, and the right column shows $t=300$. \label{fig:map}}
\end{figure}

The transition between end states with single and double flux tubes occurs at a critical $\alpha$ between $0.01$ and $0.05$. In our case, the region of turbulence coincides with the region of field line mixing, namely the kidney-shaped region best seen in the colour maps of Figure \ref{fig:map}. The transition in the final state is triggered when a particular hyperbolic (index $-1$) periodic orbit in the initial state moves inside the mixing region. This hyperbolic orbit is located well outside the mixing region at $(x,y) \approx (-4.5, 0)$ when $\alpha=0.001$ (Figure \ref{fig:map}a), moves closer [at $(x,y)\approx(-3.8,0.06)$] for $\alpha=0.01$ (Figure \ref{fig:map}d), and is eventually inside the mixing region for $\alpha=0.05$ (Figure \ref{fig:map}g). This changes the topological degree of the turbulent region from 2 to 1.

Notice that there is an asymmetry in the two tubes produced by the turbulent relaxation, and this asymmetry increases as $\alpha$ is increased. This is seen by comparing panels (e) and (f) in Figure \ref{fig:alpha} with panels (b) and (c). Firstly, the separating motion of the tubes in the $x$-direction is influenced by the background field. (If there were no background field, the tubes would simply move apart symmetrically about $x=0$). Note that we have repeated the simulation with a larger domain in $x$ with identical results at $t=300$, confirming that the background field causes the asymmetry, rather than the numerical boundary conditions. Secondly, the pattern of reversed-sign $\bar{\lambda}$ around each tube is different. Owing to the direction of rotation of ${\bf B}_\alpha$ with respect to the two tubes, there is a more significant current sheet outside the left-hand tube than outside the right-hand tube, seen clearly for $\alpha=0.01$. For $\alpha=0.001$, the background field is too weak to produce noticeable asymmetries.

As $\alpha$ is increased further beyond $0.01$, the separation of the two tubes becomes so small that the left-hand tube is eventually engulfed by the right-hand tube. The run with $\alpha=0.05$ is interesting because it is just past the transition point between double and single tube final states. In this run, the initial turbulent relaxation leaves a vestige of the second tube at $t=100$ (Figure \ref{fig:alpha}h), with a strong current sheet outside it. This current sheet is sharp enough that it undergoes resistive decay by time $t=300$, removing the second tube. From this, we see that the precise location of the transition point between asymptotic states with one and two tubes is likely to be dependent on the resistivity. On the other hand, the nature of the final state of the turbulent relaxation (\textit{e.g.} at $t=100$) is conjectured to be independent of the resistivity.

\section{Discussion}

This numerical experiment shows that one must choose the boundary appropriately if one is to correctly predict the end-state topology based on the topological degree of the initial state. The practical application of such a prediction is therefore dependent on being able to predict the extent of the turbulent relaxation sufficiently accurately. In our case, the region of turbulent relaxation is largely determined by pre-existing mapping complexity in the initial magnetic field. Therefore one makes the correct prediction by considering the chaotic mixing region of the colour maps in the initial states (Figure \ref{fig:map}).

In other situations, it may be difficult to predict the extent of the turbulent region before the onset of dynamical relaxation. For example, Bareford \textit{et al.} \cite{2013A&A...550A..40B} began with a laminar magnetic field structure not containing current sheets. Only once the kink instability had led to the onset of a turbulent relaxation did it become clear that the extent of the turbulent region would be about 1.8 times the diameter of the initial loop. It was suggested by Bareford \textit{et al.} \cite{2013A&A...550A..40B} that, due to the presence of zero net vertical current, their relaxation region was more localised than previous simulations by Browning \textit{et al.} \cite{2008A&A...485..837B} in which turbulence filled the whole domain. However, our simulations with a net vertical current still have a localised turbulent region (\textit{e.g.}, the $\alpha=0.1$ case presented here, or the ``$S^3$'' case described by Wilmot-Smith \textit{et al.} \cite{2011A&A...536A..67W}).

From a practical point of view, it is very desirable to predict not only the topology (\textit{e.g.}, number of flux tubes) of the end state, but also the amount of magnetic energy released. A possible approach is to apply Taylor theory---assuming conservation of total magnetic helicity---restricted to the turbulent region \cite{2013A&A...550A..40B}. This would predict a linear force-free field within that region. For our $\alpha=0.1$ simulation, the field does relax to a much smoother and symmetric spatial distribution of $\lambda$. But, according to the topological degree, the cases $\alpha=0.01$ and $\alpha=0.001$ cannot relax to the Taylor state, and indeed this is what our simulations show. We see the formation of two separate flux tubes of oppositely-signed $\lambda$. However, even in the case where the topological degree is consistent with a Taylor state, we find that the resulting flux tube is surrounded by a region of oppositely signed $\lambda$, such that a field with constant (or piecewise-constant) $\lambda$ is not clearly appropriate.

The physical nature of the degree constraint is nothing more or less than the freezing-in of the magnetic topology on the side boundaries of the turbulent (non-ideal) region. This constraint will exist whenever the turbulent region is localised within a wider ideal region. In our parameter study, the transition between final states with one and two flux tubes may be thought of as a change in the dominance of the contribution to the field line mapping from ${\bf B}_{\rm braid}$ compared with ${\bf B}_\alpha$. But ultimately it is the initial degree of the mapping restricted to the turbulent region that constrains the evolution.

Our assumption of a periodic domain is inessential. Although the results here are presented for the case of periodic $z$-boundaries, we have repeated the simulations for line-tied $z$-boundaries (${\bf v}=0$), as would be appropriate for the fast relaxation of coronal loops in the solar atmosphere. The qualitative finding of a transition between double and single tube final states as $\alpha$ is increased remains valid. The main difference is that the two tubes for $\alpha=0.001$ and $\alpha=0.01$ are restricted from moving apart by the line-tying of their magnetic footpoints.

Finally, we note that, although we have illustrated with resistive-MHD simulations, the degree constraint is purely a property of the global magnetic field. It is applicable more generally, relying neither on the fluid approximation nor any particular physics assumed within individual reconnection sites.

%\disclaimer{Insert disclaimer text here.}

%\ethics{Insert ethics text here.}

\acknowledgments

{ARY drafted the manuscript. AJBR and ARY ran the numerical simulations and generated the output. All authors contributed to design of the study and analysis of the results, and gave final approval for publication.}

{The Lare3D source code needed to replicate our calculations is available from the CCPForge repository \url{http://ccpforge.cse.rl.ac.uk/gf/}.}

{Numerical simulations used the UKMHD cluster at the University of St Andrews, funded by STFC and SRIF. We thank the referees for useful suggestions.}

{The authors were supported by STFC consortium grant ST/K000993/1 and ST/K001043/1 to the universities of Dundee and Durham.}

{We have no competing interests.}

%%%%%%%%%% Insert bibliography here %%%%%%%%%%%%%%
\bibliographystyle{rspublicnat}
\bibliography{lundquist}

\end{document}